\input{aipcheck}

\documentclass[
    ,final            
  ]
  {aipproc}

\layoutstyle{6x9}

\begin{document}

\title{Rotochemical heating in millisecond pulsars with Cooper pairing}


\author{Crist\'obal Petrovich}{
address={Departamento de Astronom\'\i a y Astrof\'\i sica, 
		Pontificia Universidad Cat\'olica de Chile,
		Casilla 306, Santiago 22, Chile.   }}

\author{Andreas Reisenegger}{
address={Departamento de Astronom\'\i a y Astrof\'\i sica, 
		Pontificia Universidad Cat\'olica de Chile,
		Casilla 306, Santiago 22, Chile.   }}

\begin{abstract}
When a rotating neutron star loses angular momentum, the reduction
in the centrifugal force makes it contract. This perturbs each fluid element,
raising the local pressure and originating deviations from beta equilibrium
that enhance the neutrino emissivity and produce thermal energy.
This mechanism is named \textit{rotochemical heating} and has previously been studied for
neutron stars of non-superfluid matter, finding that they reach a
quasi-steady state in which the rate that the
spin-down modifies the equilibrium concentrations is the same to that 
of the neutrino reactions restoring the equilibrium. 
On the other hand, the neutron star interior is believed to contain
superfluid nucleons, which affect the thermal evolution of the star
by suppressing the neutrino reactions and the specific heat, and opening new 
Cooper pairing reactions.

In this work we describe the thermal effects of Cooper pairing 
with spatially uniform energy gaps of neutrons and protons
on rotochemical heating in millisecond pulsars (MSPs) when only 
modified Urca reactions are allowed. We find that the chemical 
imbalances grow up to a value close to the energy gaps, which is higher
than the one of the nonsuperfluid case. Therefore, the surface temperatures
predicted with Cooper pairing are higher and explain the recent
measurement of MSP J0437-4715.
\end{abstract}

\maketitle


\section{Introduction}

The main motivation to study the thermal evolution of neutron stars
is that contrasting theoretical predictions with 
the thermal emission measured from neutron stars (NSs) has
the potential to provide constraints 
on their inner structure. In the existing literature,
several detailed cooling calculations have been compared 
to the few estimates available for the surface temperatures 
of neutron stars (see \cite{yak04} for a 
review and references). These calculations are based
on the early passive cooling, which is at first neutrino-dominated.
On the contrary, we focused our study on the late thermal
evolution, where the cooling is driven by photon emission at 
ages greater than $\sim10^5$yr.

\subsection{Rotochemical heating}

Several mechanisms can keep NSs hot beyond the standard cooling
timescale $\sim10^7$yr, among them \textit{rotochemical heating}.
The latter was first proposed in \cite{reis95} 
and then improved in \cite{FR05} by considering 
the internal structure of non-superfluid neutron stars 
via realistic equations of state (EOSs) in the framework of general relativity.
It works as follows. The reduction of the centrifugal force makes the NS contract. 
This perturbs each fluid element, raising the local pressure and originating deviations 
from beta equilibrium, which are quantified by the chemical imbalances 
$\eta_{npl}=\mu_n-\mu_p-\mu_l$, where $n$, $p$, are neutrons and protons
respectively, and  $l$ stands for leptons (electrons and muons). 
On the other hand, the neutrino reactions 
tend to restore the beta equilibrium, being more efficient as $\eta_{npl}$ 
grows. In this sense, the evolution equations for the chemical imbalances have the 
following form:
\begin{eqnarray}
\label{eq:evolucion_eta1}
\dot{\eta}^\infty_{npl} &=&  -Z_{npe}\Delta\tilde{\Gamma}_{npe}
- Z_{np}\Delta\tilde{\Gamma}_{np\mu} + 2W_{npl}\Omega \dot{\Omega}
\end{eqnarray}
where the terms $Z_{np}$, $Z_{npe}$, $Z_{np\mu}$, $W_{npe}$, and $W_{np\mu}$
are constants that depend on the stellar structure, and $\Omega \dot{\Omega}$
is the product of the angular velocity and its time
derivative (proportional to the spin-down power). Additionally, we have
introduced the net reaction rate integrated over the core, defined as
$\Delta\tilde{\Gamma}_{npl}=\tilde{\Gamma}_{n\rightarrow pl}-\tilde{\Gamma}_{pl\rightarrow n}$.

The evolution of the temperature of the isothermal interior, redshifted to a 
distant observer, $T_\infty$, is given by the thermal 
balance equation \cite{thorne77} 
\begin{eqnarray}
\label{eq:evolucion_Ti}
\dot{T}_\infty & =& \frac{1}{C} \left( L_H^{\infty}-L_\nu^\infty - L_\gamma^\infty\right),
\end{eqnarray}
where $C$ is the total heat capacity of the star, $L_\nu^\infty$ is the total 
power emitted as neutrinos due to Urca reactions,
and $L_\gamma^\infty$ is the power released as thermal photons. 
The heating term $L_H^{\infty}$  produced by each Urca-type reaction 
is defined as
$L_H^{\infty}= \eta_{npe}^{\infty}\Delta\tilde{\Gamma}_{npe}+\eta_{np\mu}^{\infty}\Delta\tilde{\Gamma}_{np\mu}$.

The most relevant feature of these equations is that, eventually, 
the system reaches a quasi-steady state where 
the rate at which spin-down modifies the equilibrium concentrations is the same at which 
neutrino reactions restore the equilibrium (see figure 1). 
This implies a conversion of rotational energy into
thermal energy and an enhanced neutrino emission originated by a departure from
the beta equilibrium. 
Thus, this mechanism can keep old millisecond pulsars (MSPs) warm,
at temperatures $\sim 10^5$ K.

\subsection{Cooper pairing}

Cooling curves usually consider the effects of nucleon 
superfluidity  on the thermal evolution of NSs.
Superfluidity is produced by Cooper pairing of baryons 
due to the attractive component of their 
strong interaction, and it is present only when the temperature $T$
of the matter falls below a critical temperature $T_c$.
However, the physics of these interactions is rather uncertain 
and very model-dependent, and so is the critical temperature obtained
from theory  (see \cite{lombardo}).
An important microscopic effect is that the onset of superfluidity
leads to the appearance of a gap $\Delta$ in the spectrum of 
excitations around the Fermi surface.
This gap in the spectrum considerably reduces the neutrino reactions and the specific heat 
involving superfluid species (neutrons and protons in the core)
\cite{yak01}, and therefore, changes the evolution of rotochemical heating. 

\begin{figure}[!h]
\includegraphics[width=12cm,height=7cm]{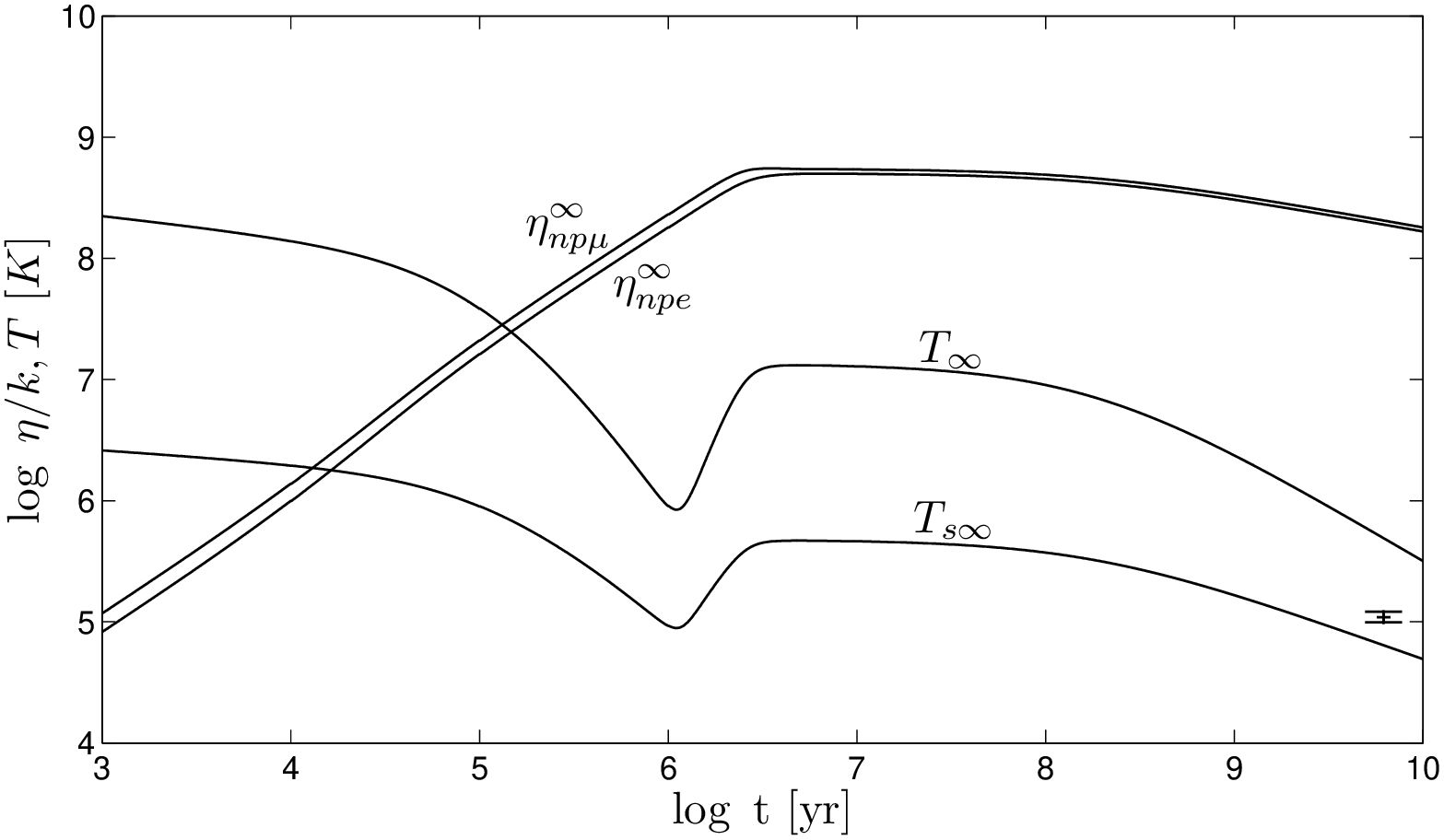}
\end{figure}

\begin{figure}[!h]
\includegraphics[width=12cm,height=7cm]{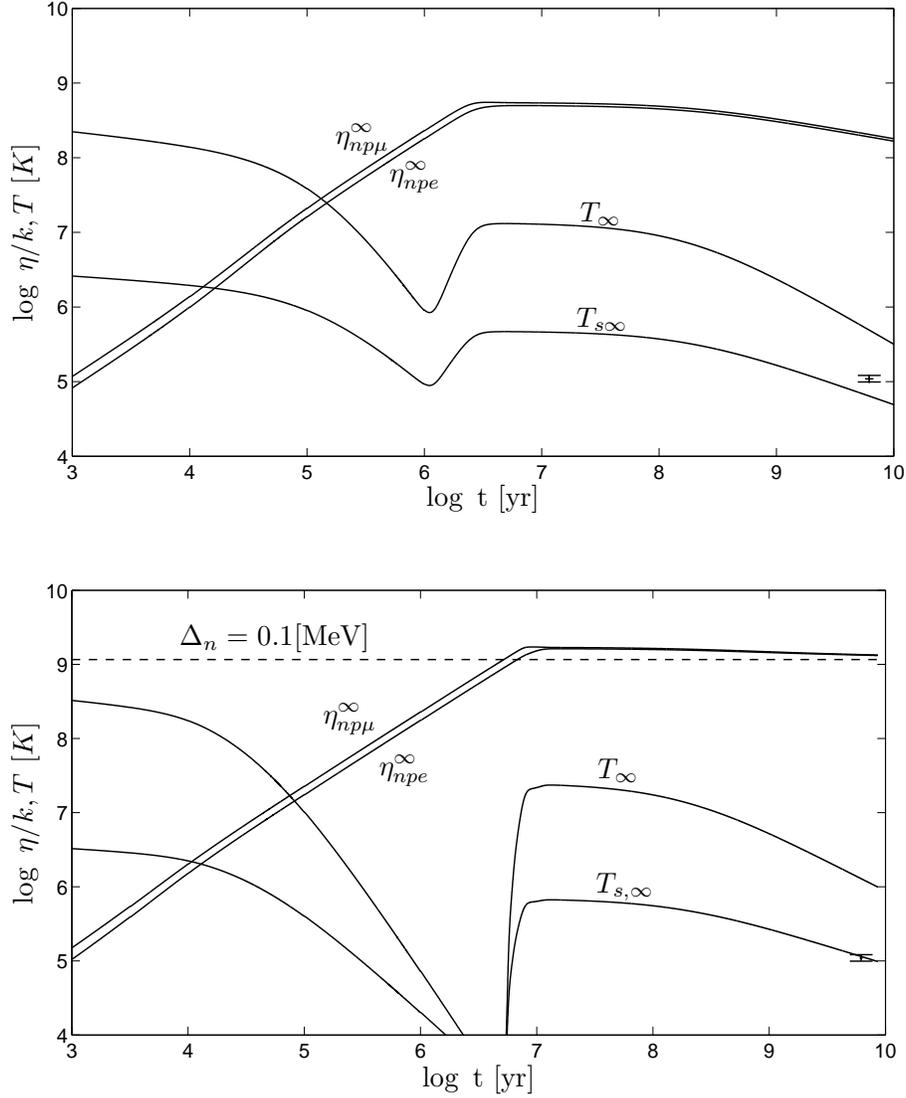}
\caption{
Evolution of the internal temperature $T_\infty$, the surface temperature $T_{s,\infty}$ and the
chemical imbalances $\eta_{npe}^\infty$, $\eta_{npu}^\infty$ for a 
star with the parameters fixed to the millisecond pulsar J0437-4715, i.e. a
mass of $1.76 M_{\odot}$ \cite{verbiest}, built with the 
A18 + $\delta\upsilon$ + UIX* EOS, and a magnetic field $B=2.8\cdot10^8$ G. 
The  initial conditions are $T_{\infty}=10^9$ K, null chemical imbalances, and
an initial period of $P_0=1$ ms. The error bar is the $90\%$ confidence level for 
the surface temperature measured for the millisecond pulsar J0437-4715 \cite{kargaltsev04} at its 
current spin-down parameters.
\textit{Upper panel:} nonsuperfluid case (null energy gaps). 
\textit{Lower panel:} superfluidity of neutrons with $\Delta_n=0.1$ MeV (dashed line).}
\end{figure}

\section{Neutrino reactions and Cooper pairing}

We consider models in which modified Urca reactions are the main 
neutrino emission mechanism:
\begin{eqnarray}
n+N_i\rightarrow p+N_f+e^-+\bar{\nu}_e \\
p+N_i+e^-\rightarrow n+N_f+\nu_e, 
\end{eqnarray}
where the subscripts $i$ and $f$ stand
for the initial and final state of the spectator nucleon $N$.

As in \cite{villain},  we  compute the net reaction rate $\Delta\tilde{\Gamma}_{npl}$
and the neutrino emissivity for these reactions numerically,
considering the chemical imbalances $\eta_{npl}$, the 
energy gaps $\Delta$ in the energy spectra of the nucleons,
and the temperature as free parameters (see details in \cite{petro09}). We find 
that at low temperatures, i.e. $T\ll \eta_{npl}$, which is the regime of interest in
rotochemical heating (see the upper panel of figure 1), these reactions
are almost completely blocked due to the energy gap  
when $\eta_{npl}<\Delta$. However, several reactions are opened
when $\eta_{npl} > \Delta$. 
This is the most important effect of Cooper
pairing in rotochemical heating since the quasi-steady state will
be reached when the restoring mechanism given by the neutrino reactions
becomes sufficiently important to conterbalance the effect of the 
spin-down forcing mechanism.

\section{Results and discusion}

Fig. 1 shows that, for the superfluid case, the reactions are blocked
until the chemical imbalances overcome 
the value of the energy gap of the neutrons $\Delta_n$, as argued above.
These chemical imbalances are higher than those 
achieved in the non-superfluid case.
This effect lengthens the timescale at which the system reaches the 
quasi-steady state, and implies that, in the presence of superfluidity, the chemical
energy is larger and dissipated later to
reheat the star. This makes it possible to fit the
observation of the MSP J0437-4715 \cite{kargaltsev04}, unlike the non-superfluid case.
Finally, for several EOS in which modified Urca reactions
are the dominant neutrino emission mechanism, this observation constrains the
energy gaps to lie in the range
$0.05 [ \mbox{MeV}]<\mbox{min}\left(  \Delta_n+3\Delta_p,3\Delta_n+\Delta_p \right)<0.45 [ \mbox{MeV}]$
\cite{petro09}.





\bibliographystyle{aipproc}   




\end{document}